\documentclass[letterpaper,conference,twocolumn]{IEEEtran}

\usepackage{color}
\usepackage{graphicx}

\ifCLASSOPTIONcompsoc 
	\usepackage[caption=false,font=normalsize,labelfont=sf,textfont=sf]{subfig}
\else
	\usepackage[caption=false,font=footnotesize]{subfig}
\fi

\usepackage[cmex10]{amsmath}
\usepackage{amssymb}
\usepackage{mathtools}
\usepackage{float}
\usepackage{tikz}
\usepackage{epstopdf}

\graphicspath{ {figures/} }

\newcommand{\saeed}[1]{}

\newcommand{\fig}[1]{Fig.~\ref{#1}}

\newcommand{\doc}[1]{}

\usetikzlibrary{positioning}
\usetikzlibrary{fit}

\tikzstyle{axis-line}=[thick, ->, >=stealth]
\tikzstyle{axis-tick}=[thin]
\tikzstyle{arrow} = [thick,->]

\tikzstyle{block}=[rectangle,black,draw, text centered, minimum height=0.7cm]
\tikzstyle{summul} = [draw, shape=circle, text centered,inner sep=0]

\title{Multicarrier PAPR Reduction by Iteratively \\Shifting and Concentrating the Probability Measure}

\date{\today}

\author{\IEEEauthorblockN{Saeed Afrasiabi-Gorgani, Gerhard Wunder}
\IEEEauthorblockA{Heisenberg Information and Communication Theory Group\\
Freie Universit\"{a}t Berlin\\
s.afrasiabigorgani@tu-berlin.de, gerhard.wunder@hhi.fraunhofer.de}}

\begin{document}

\maketitle

\begin{abstract}
The peak power problem in multicarrier waveforms is well-known and imposes substantial limitations on wireless communications. As the quest for investigation of enabling technologies for the next generation of wireless communication systems 5G is at its peak, the problem  is re-emerging in a much broader range of technologies. However, despite numerous publications on the topic, there is no well-established structure available for the problem, which motivates a boost in research. In this paper, a novel peak power reduction algorithm is proposed which offers a substantial Peak-to-Average Power Ratio (PAPR) reduction and exhibits high potential for further refinements. Mathematical tractability of the algorithm is expected to be of particular importance to this end. A remarkable early observation is a PAPR reduction of about 4.5 dB for 64 subcarriers with a rate loss of 0.5 bits per complex data symbol in an OFDM scheme, which is half the rate loss that other methods require in this class of algorithms.
\end{abstract}




\section{Introduction}




\saeed{why different choices of a say half subset of signs make a difference? I remember it did! but what I suggest in the paper is that when there are many signs left the impact low, or a small subset is better. But what about distribution and location of the signs? how and why it makes a difference?}

The PAPR problem has been an active research topic since Orthogonal Frequency Division Multiplexing (OFDM) became the popular waveform in wireless communications. High PAPR is an inherent characteristic of multicarrier waveforms and requires a power amplifier with a larger linear range to avoid the distortion which causes performance degradation or in-band distortion and more crucially, out-of-band radiation. 
 Therefore, either a back-off is necessary or the PAPR problem needs to be handled in baseband processing of the waveform. Back-off refers to reducing the average power of the transmit signal so that the peaks fall less frequently in the nonlinear range.
 In other words, the power amplifier needs to be designed such that enough linear range is available for the transmit power levels of interest.
 This way of dealing with the high PAPR problem causes reduced amplifier efficiency. In general, the PAPR problem is more critical in the uplink due to limited battery life of portable devices. A famous example is the choice of Single Carrier Frequency Division Multiple Access (SC-FDMA) for uplink in LTE standards, where a low-PAPR waveform had to be used instead of a reduction method. However, it must be mentioned that energy efficiency in basestations, i.e. in downlink direction, is an important concern as well \cite{EnergyAware2010}.
 
As new technologies emerge, particularly in context of 5G, the limitations imposed by the PAPR problem remain substantial. For instance, millimeter-wave (mmWave) technology is one of the main candidates for providing substantially higher data rates promised by 5G. The current technology for RF front-ends is reported to offer lower efficiency in power amplifiers operating in mmWave frequency bands \cite{mmWaveSamsung2011}, which makes the use of multicarrier waveforms limited due to the PAPR problem. 
The other viewpoint is energy efficiency of communication networks. From the overall energy consumption perspective, the power efficiency in basestations is critical, while a major part of which is due to high PAPR \cite{EnergyAware2010}.

Numerous methods have been proposed to either reduce the PAPR of the OFDM signal or to control the added distortion from the nonlinear path by transmitter- and/or receiver-side processing. There are roughly two categories: \cite{1421929} a) The distortion-based methods that perform deliberate clipping to control the PAPR, which use different sorts of processing to keep the out-of-band radiation and the performance degradation limited. b) The distortionless methods that modify the signal to reduce the PAPR.

A group of distortionless methods 
  rely on manipulating the data points such that they remain in the set of constellation points of the designed system and the PAPR of the signal is reduced. A simple way to perform this method is to rotate the phases of the complex data symbols by a  number of phase vectors and generate the corresponding signal segments. Then the candidate signal which has the least PAPR is chosen. This method is referred to as Selected Mapping (SLM) \cite{543811}. If the possible phase rotations are limited to $\{1,-1\}$, it is called sign selection.

The proposed method in this paper falls in a line of research whose objective is to 
gain as much PAPR reduction as possible against limited rate loss at the cost of high computational complexity.  This objective is being followed by different approaches, an appropriate account of which is beyond the scope of this paper. The reader is referred to \cite{gerhard2013SPM} for a fundamental study of the PAPR problem.  Therefore, we narrow down our attention to a subclass of PAPR reduction methods which focus on distortionless constant-power methods that solely use sign selection.
Two of such methods which provide high PAPR reduction are: derandomized algorithm \cite{Sharif2004constantPMEPR,Afrasiabi2015derandomized} and a greedy algorithm proposed in \cite{Sharif2009sign}. In both cases, an upper bound on the worst-case PAPR is derived.

The algorithm proposed in this work provides PAPR reduction by choosing sign of complex data symbols one by one. By each sign decision, the goal is to reduce the expected value of the PAPR random variable conditioned on the already fixed signs, as a function of remaining random sign variables and fixed data symbols. The process can be described by considering the probability measure of PAPR, which is concentrated around its expected value \cite{Sason2011-concentration}. Each step of the algorithm shifts the probability measure of the resulting PAPR to left, which has less randomness due to fixed signs. Therefore, the probability measure of the updated PAPR has a stronger concentration and is shifted to left until the last step, where the PAPR is no more random and is equal to the last expected value.

The algorithm is presented for OFDM waveform and PAPR metric. However, as waveform design is an active area of research in direction of 5G \cite{gerhard5G}, applicability and performance of the method for other waveforms is a concern. A simplifying characteristic of OFDM waveform is isolation of the segments of the signal that correspond to separate blocks of data symbols. This simplicity is exploited in this paper  to develop the main body of the algorithm. Extension to other waveforms should be a direction for further research. Furthermore, alternative metrics are suggested, for instance the Cubic Metric \cite{CMmotorola}, that are reported to be superior to PAPR in determining the required power back-off. The proposed algorithm is readily applicable to any metric. However, analytical evaluation of the performance needs a separate treatment.

The paper is organized as follows: Section~\ref{sec:pre} includes the basic definitions and signal model. Section~\ref{sec:algorithm} describes the proposed algorithm. The PAPR reduction performance of the algorithm is analyzed in Section~\ref{sec:analysis}. The performance considering different parameters of the algorithm is discussed in Section~\ref{sec:performance} based on computer simulations. Finally, Section~\ref{sec:conclusion} concludes the paper.

\emph{Notation:} Capital letters are used to denote random variables. Vectors are distinguished by bold face letters. Consecutive elements of $\mathbf{x}$ are shown by $x_{n:m}$. Probability distribution function of $X$ is denoted by $\mathrm{P_X}$, where $X$ is dropped if not necessary. Finally, $|\mathcal{S}|$ denotes cardinality of the set $\mathcal{S}$.

\section{Preliminaries}
\label{sec:pre}

Let $\mathcal{M}$ denote the set of constellation points and $n$ be the number of subcarriers in the OFDM scheme. It is assumed that for all $c\in\mathcal{M}$, $-c\in\mathcal{M}$ as well. Consider the OFDM baseband signal model
\begin{equation}
	s(t)=\frac{1}{\sqrt{n}}\sum_{k=0}^{n-1} C_k e^{j 2\pi k t/\tau} \quad t\in[0,\tau),
\end{equation}
where $C_k\in \mathcal{M}$ and $\tau$ is the OFDM symbol duration. This segment of $s(t)$ corresponds to a single block of $n$ complex data symbols.\saeed{revise} The OFDM symbols do not overlap in time, hence a PAPR reduction algorithm can be performed without considering the past or future blocks of data symbols. 

The Peak-to-Average Power Ratio (PAPR) is a random variable and is defined over one OFDM symbol as
\begin{equation}
\mathrm{PAPR}(\mathbf{C}) = \frac{1}{p_a} \max_{t\in[0,\tau)} |s(t)|^2,
\end{equation}
where $p_a$ is the average power of $s(t)$ and \saeed{sure?}
\begin{equation}
	p_a=\mathrm{E}[|s(t)|^2]=\mathrm{E}[|C_i|^2].
\end{equation}
Crest Factor (CF) is also a commonly used metric which is essentially the same as PAPR and is defined as $\mathrm{CF}=\sqrt{\mathrm{PAPR}}$. This characteristic of a signal is often represented by the Complementary Cumulative Distribution Function (CCDF) of PAPR or CF. For a consistent discussion on the reduction performance, consider \emph{effective PAPR} defined as the PAPR value for which the CCDF equals~$10^{-3}$. In words, it is the value that PAPR exceeds with a probability of~$10^{-3}$.

Consider the i.\@i.\@d random vector $\mathbf{X}$ of sign variables, i.e., $X_k\in\{-1,1\}$, with $\mathbb{P}(X_i=1)=0.5$. An OFDM symbol generated from point-wise multiplication of the sign variables and the complex data symbols can be written as
\begin{equation}
	s(t,\mathbf{X},\mathbf{C})=\frac{1}{\sqrt{n}}\sum_{k=0}^{n-1} X_k C_k e^{j 2\pi k t/\tau}\quad t\in[0,\tau).
\end{equation}
In a sign selection method, the sign variables are chosen such that the PAPR is reduced. Obviously, a trivial solution is exhaustive search over $2^{n-1}$ possible combinations. Note that negation of a signal segment does not alter its PAPR, hence the first sign variable can be fixed to 1.

\section{The proposed algorithm}
\label{sec:algorithm}

Consider CF written as 
\begin{equation}
	f(\mathbf{x},\mathbf{c})=\frac{1}{\sqrt{p_a}}\max_{t\in[0,\tau)} |s(t,\mathbf{X}=\mathbf{x},\mathbf{C}=\mathbf{c})|.
\end{equation}
In order to save space, the condition that a random variable takes a constant value will be written as, e.g., $x_1$ instead of $X_1=x_1$. In the proposed algorithm, for a given block of data symbols $\mathbf{c}$, the sign selection for $x_j$ is performed such that
\begin{equation}
\mathrm{E} [f(\mathbf{X},\!\mathbf{C})| x^\ast_{0:j}, \mathbf{c}]\!\!=\!\min_{x_j\in \{-1,1\}} \mathrm{E} [f(\mathbf{X},\!\mathbf{C})|x^\ast_{0:j-1},x_j,\mathbf{c}],
\label{eq:sign-selection-criterion}
\end{equation}
i.e., the sign which gives the lower expected value is chosen. Notice that $.^\ast$ distinguishes the decided signs.
This process yields a sequence of non-random quantities
\begin{align}
	z_0&=\mathrm{E}_{X_{0:n-1}} [f(\mathbf{X},\mathbf{C})|\mathbf{c}] \nonumber \\
	&\vdots\nonumber \\
	z_j&=\mathrm{E}_{X_{j:n-1}} [f(\mathbf{X},\mathbf{C})|x^\ast_{0:j-1},\mathbf{c}] \nonumber\\
	&\vdots\nonumber\\
	z_{n}&=\mathrm{E}_\emptyset [f(\mathbf{X},\mathbf{C})|\mathbf{x}^\ast,\mathbf{c}]=f(\mathbf{x}^\ast,\mathbf{c}).
	\label{eq:z-expectations}
\end{align}
The first conditional expectation conditioned on the random vector $\mathbf{C}$, denoted by $Z_0$, is referred to as a \emph{partial expectation} in this paper. The term partial is used in relation to the \emph{full} expectation $\mu=\mathrm{E} [f(\mathbf{X},\mathbf{C})]$, which is a constant value. An essential part of presenting the algorithm is postponed to Section~\ref{sec:analysis}, where it will be shown how this choice of signs yields reduction in CF, or equivalently PAPR.

The proposed algorithm is essentially a local search with conditional expectation as the cost function. Derivation of a closed-form expression to evaluate \eqref{eq:sign-selection-criterion} appears to be complicated for CF. More promising approaches could be using bounds or alternative metrics for $f(\mathbf{X},\mathbf{C})$. However, it will be shown by simulations that using estimates of conditional expectations provides sufficient accuracy to achieve desirable performance. Therefore, we take estimation as an immediately available tool to remove the complexity and develop the algorithm. The motivation for this choice of cost function is considerable PAPR reduction performance, as shown in Section \ref{sec:performance}, and mathematical tractability. As explained later, the latter is based on true values of the conditional expectations. 
However, it will be seen that even a rather coarse estimation provides enough accuracy to benefit from the insights gained from analysis.

The expected value can be estimated by the so-called sample average \saeed{1/q?}
\begin{equation}
\mathrm{\hat{E}} [f(\mathbf{X},\mathbf{C})|x_{0:j-1},\mathbf{c}]=\sum_{i=1}^q f(\mathbf{x}_i,\mathbf{c}),
\end{equation}
which uses $q$ \emph{shots}, i.e., $q$ realizations of $X_{j:n-1}$. By standard analysis\saeed{cite}, sample average is unbiased\saeed{define?} and its variance is $\sigma_f/q$, where $\sigma_f$ is variance of $f(\mathbf{X},\mathbf{C})$. Further analysis is beyond the scope of this paper. Hence, we shall suffice to the fact that the variance of the estimator decreases for higher $q$.

\subsection{Refined algorithm}

\begin{figure}[t]
	\centering
	\includegraphics[width=\columnwidth]{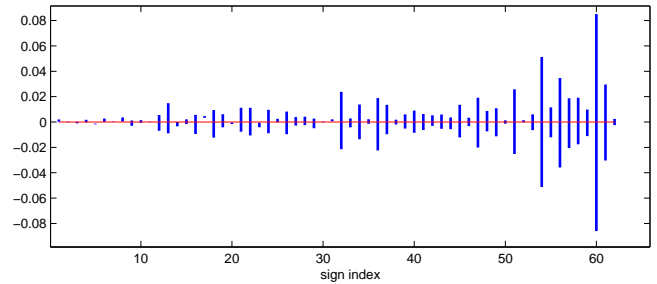}
	\vspace{-0.7cm}
	\caption{A snapshot of the estimated differences $z^+_j-z_j$ and $z^-_j-z_j$ normalized by $z_j$ for $n=64$ and 16-QAM. Vertical lines connect the two difference values for each sign decision.\doc{not normalized! done by v00-test2}}
	\label{fig:plus-minus-diff-64-50000shots}
		\vspace{-0.4cm}
\end{figure}

The sequence of conditional expectations in \eqref{eq:z-expectations} are essentially an averaging over the values that $f$ takes for possible realizations of fixed and random sign variables. The number of values over which the averaging is carried out decreases for the sequence elements with higher indices, i.e., as the algorithm proceeds with further sign choices. Therefore, intuition points out that each function value gains a higher contribution to the average value. It suggests that the conditional expectations
\begin{align}
z^\pm_j&=\mathrm{E} [f(\mathbf{X},\mathbf{C})|x^\ast_{0:j-1},X_j=\pm 1,\mathbf{c}],
\end{align}
which need to be estimated for making decision on $x^\ast_j$, are likely to have a larger difference with $z_j$ as $j$ increases. For generously high $q$, \fig{fig:plus-minus-diff-64-50000shots} shows a snapshot of the differences $z^+_j-z_j$ and $z^-_j-z_j$.\saeed{revise}

This intuition motivates a refined version of the algorithm in the sense that it uses a subset of sign variables for PAPR reduction which have higher impact. Let $m$, such that $1\leq m < n$, be the index of the first sign variable. The refined algorithm uses only $x_{m:n-1}$, i.e., the last $n-m$ sign variables. The first $m$ sign variables are left undecided, i.e., fixed to 1.


\subsection{Side Information}
The distortionless PAPR reduction methods that modify the data symbols, for instance by phase rotation, need to undo this operation in the receiver. The information required to perform detection, which is often the exact sequence of the modifications, is called Side Information (SI). This requirement has been typically handled either by assuming feasibility of a reliable transmission of the SI in parallel or by modifying the algorithm to avoid explicit SI. The former approach poses complicated issues: timely reception of SI and possible latency, and existence or reliability of such parallel channels. 
The latter approach is a different algorithmic challenge and a major subproblem, which stops a systematic research on the PAPR problem. 

For the proposed algorithm, and those using sign selection, detection could be done by discarding the sign of the data symbols. Note that this causes the same rate loss as the amount of SI required for undoing the process when all or a pre-determined subset of data symbols undergo a modification. This is clearly a preferable approach. Given this view, the authors suffice to only deriving the required amount of SI in the proposed algorithm, which is referred to here as rate loss.

The rate loss can be represented by 
\begin{equation}
r_l=\log_2 |\mathcal{M}| - \frac{1}{n} \log_2 \left(\frac{|\mathcal{M}|}{2}\right)^n
\end{equation}
bits per complex data symbol (b/sym). Discarding the signs indicates that only half of the constellation points, i.e., $|\mathcal{M}|/2$ of them, are used for data transmission. Therefore, as expected, the algorithm causes a rate loss of 1 b/sym. For the refined version of the algorithm, only the sign of data symbols $m$ to $n$ need to be discarded. That is, it causes a rate loss of 
\begin{align}
r_l(m)&=\log_2 |\mathcal{M}| - \frac{1}{n} \log_2 \left[ |\mathcal{M}|^{m} \left(\frac{|\mathcal{M}|}{2}\right)^{n-m} \right] \nonumber \\
&=\frac{n-m}{n}.
\end{align}
Therefore, for the case of $m=n/2$, where the performance degradation will be seen to be surprisingly small, the rate loss is 0.5 b/sym instead of 1 b/sym.

The number of bits per complex data symbol depends on the modulation order of each subcarrier, i.e., the constellation size. On the other hand, the rate loss imposed by the algorithm is fixed in terms of bits per symbol. Therefore, the overall rate loss decreases for larger constellations.

\section{Analysis of Partial Expectations}
\label{sec:analysis}

In order to analyze the behavior of the conditional expectations, consider that
\begin{align}
	\mathrm{E} &[f(\mathbf{X},\mathbf{C})|x^\ast_{0:j-1},\mathbf{c}]=\nonumber \\
	&\quad\sum_{x_{j:n-1}}\!\!\! f(x^\ast_{0:j-1},x_{j:n-1},\mathbf{c}) \mathrm{P} (x_{j:n-1}),
\end{align}
since $\mathbf{X}$ is independent of $\mathbf{C}$ and i.\@i.\@d. We have
\begin{align}
	\mathrm{E} [&f(\mathbf{X},\mathbf{C})|x^\ast_{0:j-1},\mathbf{c}]=\!\!\!\!\sum_{x_{j:n-1}} \!\!\!f(x^\ast_{0:j-1},x_{j:n-1},\mathbf{c}) \mathrm{P}(x_{j:n-1}) \nonumber\\
	&=\!\!\!\!\!\sum_{x_{j+1:n-1}} \!\!\!\!\!\Big[f(x^\ast_{0:j-1},1,x_{j+1:n-1},\mathbf{c}) \mathrm{P} (x_{j+1:n-1}) \mathrm{P}_{X_j}(1) \nonumber\\
	&\quad+ f(x^\ast_{0:j-1},-1,x_{j+1:n-1},\mathbf{c}) \mathrm{P} (x_{j+1:n-1}) \mathrm{P}_{X_j}(-1)\Big]\nonumber\\
	&=\frac{1}{2} \left( \mathrm{E} [f(\mathbf{X},\mathbf{C})|x^\ast_{0:j-1},X_j=1] \right. \nonumber\\
	&\quad\quad+\left. \mathrm{E} [f(\mathbf{X},\mathbf{C})|x^\ast_{0:j-1},X_j=-1] \right),
\end{align}
which implies that the conditional expectations in \eqref{eq:z-expectations} satisfy 
\begin{equation}
	z_0 \geq z_1 \geq \ldots \geq z_n.
	\label{eq:decreasing-trend}
\end{equation}
That is, there is a non-increasing trend in the conditional expectations starting from $z_0$ with no sign fixed to $z_n$ with all signs decided. Considering that the right-most expectation is $f(\mathbf{x}^\ast,\mathbf{c})$ itself, \eqref{eq:decreasing-trend} gives $z_0$ as an upper bound on the CF. Therefore, PAPR reduction capability of the algorithm can be analyzed by investigating the partial expectation $Z_0$, which is the \emph{random upper bound} on the reduced CF. To this end, an analytic approach using concentration inequalities and a numerical approach using estimated distribution of $Z_0$ are presented in the remainder of this section.

\subsection{Concentration of Partial Expectation}

In \cite{Sason2011-concentration}, several concentration inequalities\saeed{a brief intro?} are considered to analyze the concentration of Crest Factor of the OFDM signal around its expected value. Here, McDiarmid's inequality is applied in a slightly different way to establish concentration of the partial expectation $Z_0$ around $\mu=\mathrm{E} [f(\mathbf{X},\mathbf{C})]$.

Given that random variables $C_i$ are independent, if it can be shown for $g(\mathbf{C})$ that
\begin{equation}
	|g(\mathbf{c})-g(\mathbf{c}')|\leq d_k
	\label{eq:McDiarmid-diff}
\end{equation}
when only the $k^\mathrm{th}$ component of $\mathbf{c}$ and $\mathbf{c}'$ disagree, McDiarmid's inequality holds and states that for every $\alpha\geq 0$
\begin{equation}
	\mathbb{P}(|g(\mathbf{C})- \mathrm{E}[g(\mathbf{C})]|\geq \alpha) \leq 2 e^{-\frac{2\alpha^2}{\sum_k d_k^2}}.
\end{equation}


\begin{figure}[ht]
\centering
	\includegraphics[width=\columnwidth]{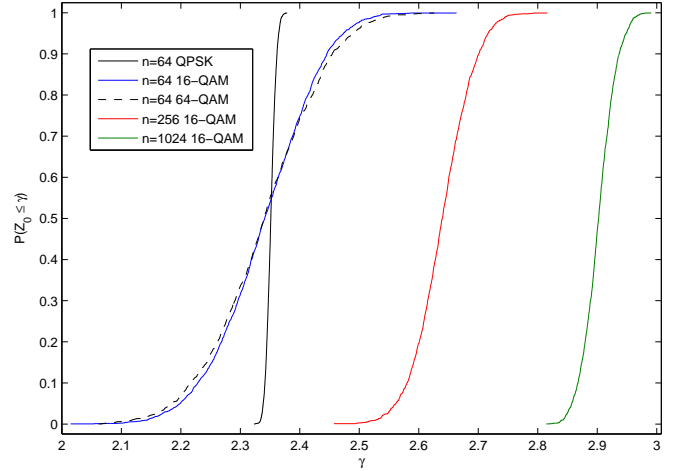}
	\vspace{-0.7cm}
	\caption{Estimated CDF for $Z_0$. \saeed{move} }
	\label{fig:estimated-CDFs}
	\vspace{-0.3cm}
\end{figure}

For our purpose, let $g$ be
\begin{equation}
	g(\mathbf{C})=\mathrm{E}_\mathbf{X} [ f(\mathbf{X},\mathbf{C})|\mathbf{C}],
\end{equation}
with expected value of
\begin{equation}
	\mathrm{E} [g(\mathbf{C})]=\mathrm{E}_{\mathbf{X},\mathbf{C}} [f(\mathbf{X},\mathbf{C})],
\end{equation}
which is the full expectation $\mu$.
The bounded differences of~\eqref{eq:McDiarmid-diff} can be shown as follows. Let
\begin{align}
	\mathbf{c}&=[c_0, c_1, \ldots, c_k, \ldots, c_{n-1}], \nonumber\\
	\mathbf{c}'&=[c_0, c_1, \ldots, c'_k, \ldots, c_{n-1}].
\end{align}
Since $|\mathrm{E}(Z)|\leq \mathrm{E}(|Z|)$, we have
\begin{align}
	|g(\mathbf{c})-g(\mathbf{c'})| &= |\mathrm{E}_\mathbf{X} [f(\mathbf{X},\mathbf{c})-f(\mathbf{X},\mathbf{c}')]| \nonumber \\
	&\leq \mathrm{E}_\mathbf{X} |f(\mathbf{X},\mathbf{c})-f(\mathbf{X},\mathbf{c}')|.
\end{align}
In addition, as $\left| \max |p(t)|\!-\! \max|q(t)|\right|\leq \max |p(t)\!-\!q(t)|$,
\begin{align}
	|f(\mathbf{X},\mathbf{c})-f(\mathbf{X},\mathbf{c}')|&\leq \frac{1}{\sqrt{p_a}} \max_{t\in[0,T)} |s(t,\mathbf{X},\mathbf{c})-s(t,\mathbf{X},\mathbf{c'})| \nonumber \\
	&=\max_{t\in[0,T)} \frac{1}{\sqrt{np_a}} |(X_k c_k-X_k c'_k) e^{j 2\pi k t/\tau}| \nonumber \\
	&\leq \frac{1}{\sqrt{np_a}} |c_k-c'_k| \leq \frac{d}{\sqrt{np_a}},
\end{align}
where $d=2\max_{c\in\mathcal{M}} |c|$ is the largest distance between constellation points in $\mathcal{M}$. As $\mathrm{E} [f(X)] \leq \mathrm{E} [g(X)]$ for $f(X)\leq g(X)$, we reach at
\begin{equation}
	|g(\mathbf{c})-g(\mathbf{c'})| \leq \frac{d}{\sqrt{np_a}}.
\end{equation}
Therefore, the McDiarmid's inequality can be written as\saeed{side}
\begin{equation}
	\mathbb{P}(\mathrm{E} [ f(\mathbf{X},\mathbf{C})|\mathbf{C}]- \mathrm{E} [f(\mathbf{X},\mathbf{C})]\geq \alpha) \leq  e^{-2\alpha^2 p_a/d^2}.
	\label{eq:McDiarmid-onesided}
\end{equation}

For instance, consider $n=64$ and 16-QAM. A good estimate of $\mu$ is 2.34, which approximately refers to a PAPR of 7.45 dB. For a probability upper-bound of $10^{-3}$, \eqref{eq:McDiarmid-onesided} can be read as: in only less than 0.001 of the cases, i.e., realizations of $\mathbf{C}$, $Z_0$ can get more than $\alpha=4.98$ away from $\mu$. In other words, the reduced PAPR can exceed at most 17.29 dB with a probability of at most $10^{-3}$. 
The bound on the reduction gain of the algorithm will be further discussed in the next section.

\subsection{A numerical approach to performance analysis}
\label{sec:num-approach}

\begin{figure}[ht]
	\centering
	\includegraphics[width=\columnwidth]{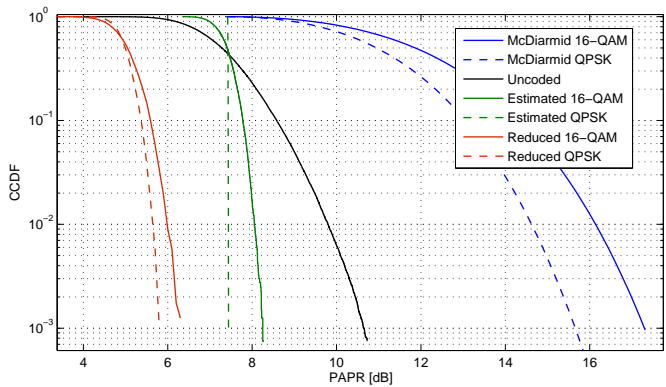}
	\vspace{-0.7cm}
	\caption{A representation of the random upper-bound on reduced PAPR by the CCDF curves, for $n=64$. }
	\label{fig:upperbound-CCDF-64}
\end{figure}

Estimated CDF curves for several number of subcarriers are shown in \fig{fig:estimated-CDFs}.
It can be seen that $Z_0$ becomes more densely concentrated about $\mu$ as $n$ increases. The other factor that affects the distribution is the modulation order.
For QPSK, $Z_0$ stays roughly equal to $\mu$. For a higher constellation size, the distribution spreads out. However, even for  high number of points, such as 256-QAM, the distribution is still fairly dense. This can be better appreciated by referring to \fig{fig:upperbound-CCDF-64} where the CCDF of $Z_0$ is used as a probabilistic upper-bound.

Due to \eqref{eq:decreasing-trend}, $\mathbb{P}(Z_0> \gamma)=p$ yields $\mathbb{P}(Z_n> \gamma)\leq p$. That is, we can obtain a CCDF representation for the upper-bound on worst-case PAPR. This is shown for $n=64$ in \fig{fig:upperbound-CCDF-64}, where the upper-bound obtained from the McDiarmid's inequality is included as well. 
Unfortunately, the concentration inequalities applied to the problem do not turn out to be very informative. On the other hand, estimated distribution of $Z_0$ provides a rather \emph{brick-wall} upper-bound.


\textbf{Remark}: Developing a mathematically tractable algorithm for PAPR reduction is not straight-forward.
Although a complete literature review is not available, it appears that in the cases where analysis of PAPR reduction capability  is possible, only a poor deterministic upper-bound on the worst-case PAPR is obtained. Although a clear relationship between the worst-case PAPR and the PAPR threshold for probability levels of interest \saeed{efficoent papr?} has not been found, such poor upper-bounds make it even more challenging. This point of view makes this specific form of upper-bounding more interesting, although it is a mixture of analysis and numerical work.


\subsection{Refined algorithm}
\saeed{idiot! just remove the extra variables from the model.}

The decreasing trend shown in \eqref{eq:decreasing-trend} clearly holds as well for any consecutive subset of the conditional expectations. Therefore, the numerical approach to the analysis of PAPR reduction capability of the algorithm can be directly applied to its refined version. That is, $Z_{m}$ becomes the random upper bound of the reduction that is done using $x_{m:n}$. Here $Z_{m}$ is the random version of $z_{m}$, similar to the relation between $Z_0$ and $z_0$.  \fig{fig:n64-pruned-bounds} shows the CCDF curves for the probabilistic upper bound for $m=n/2$ and $3n/4$ with $n=64$. It can be seen that the distribution spreads out as $m$ increases, i.e., as $Z_{m}$ is an expectation over a smaller subset of $\mathbf{X}$. Note that $Z_{m}$ is taken over $X_{0:m-1}$.

\begin{figure}[ht]
	\centering
	\includegraphics[width=\columnwidth]{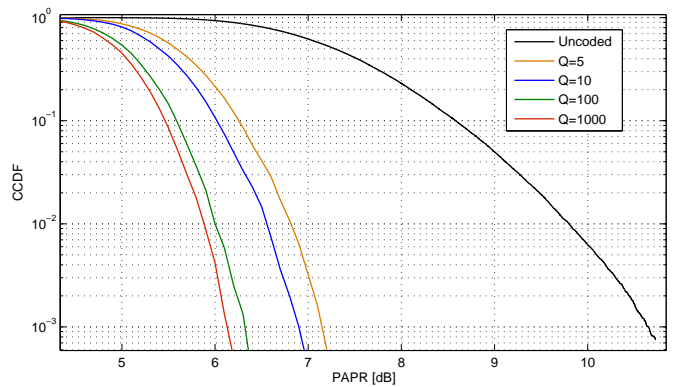}
	\vspace{-0.7cm}
	\caption{PAPR reduction for $n=64$ subcarriers for several values of $q$ and 16-QAM.}
	\label{fig:64-reduction}
\end{figure}

\section{Performance Evaluation}
\label{sec:performance}

Although development of the algorithm is done on continuous-time signal, implementation in base-band digital domain or in computer simulations is done using the discrete-time signals. Oversampling is essential to reliable measurement of PAPR, especially if it is part of an iterative algorithm. The rule-of-thumb is an oversampling factor of $L\geq4$.  Concerning the simulation setup, it should be noted that the proposed algorithm is distortionless, i.e., the error rate performance of the system is unaffected. Therefore, the simulation is done only for of transmit signal generation.

To investigate effect of estimation accuracy, determined by the number of shots $q$ used in estimation of conditional expectations, the PAPR reduction for $n=64$ subcarriers and 16-QAM for several values of $q$ is shown in \fig{fig:64-reduction}. Clearly, for higher $q$, probability of making a mistake in sign selection at each step decreases. Note that inaccuracies in absolute values of the intermediate expectations are not important as long as the error does not change their true order when calculated for +1 and -1 signs. It can be observed that even $q=5$ provides a good performance of 3~dB reduction. Note that the PAPR reductions reported in this section refer to the reduction in the effective PAPR as defined in Section~\ref{sec:pre}. For the rest of the simulations, $q=100$ is used.

\begin{figure}[t]
	\centering
	\includegraphics[width=\columnwidth]{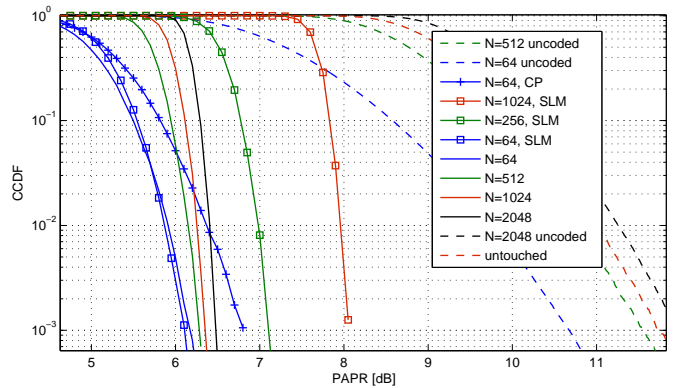}
	\vspace{-0.7cm}
	\caption{PAPR reduction achieved by the proposed algorithm for several number of subcarriers $n$, 16-QAM and $q=100$ shots.}
	\label{fig:differentN-reduction}
	\vspace{-0.5cm}
\end{figure}

The PAPR reduction performance of the algorithm for different number of subcarriers is shown in \fig{fig:differentN-reduction}. A full search, i.e. $m=1$, is assumed unless $m$ is explicitly determined. The reduction is considerable: 4.6~dB, 4.6~dB and 4.7~dB for $n=64$, $256$ and $512$, respectively. As a comparison, the derandomized algorithm as applied in \cite{Sharif2004constantPMEPR} provides a reduction of about 4.2 dB for $n=64$ and 16-QAM. \saeed{say it gets better} The greedy algorithm suggested in \cite{Sharif2009sign} offers about the same performance as that of the derandomized algorithm. 

\begin{figure}[ht]
	\centering
	\includegraphics[width=\columnwidth]{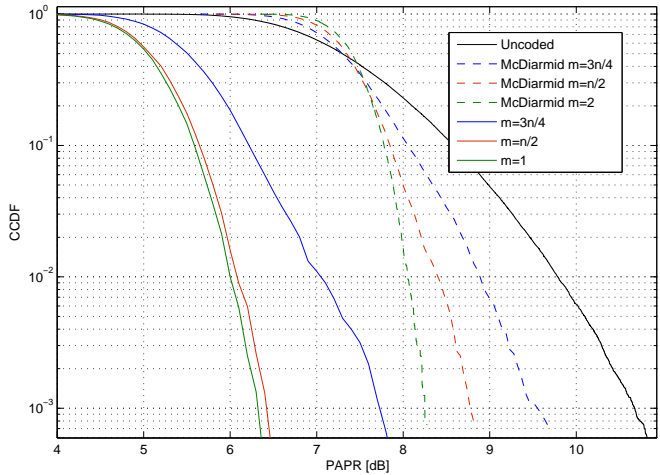}
	\vspace{-0.7cm}
	\caption{Probabilistic upper bound for PAPR reduction of the refined algorithm for $n=64$, 16-QAM and $q=100$.}
	\label{fig:n64-pruned-bounds}
\end{figure}

\begin{figure}[ht]
	\centering
	\includegraphics[width=\columnwidth]{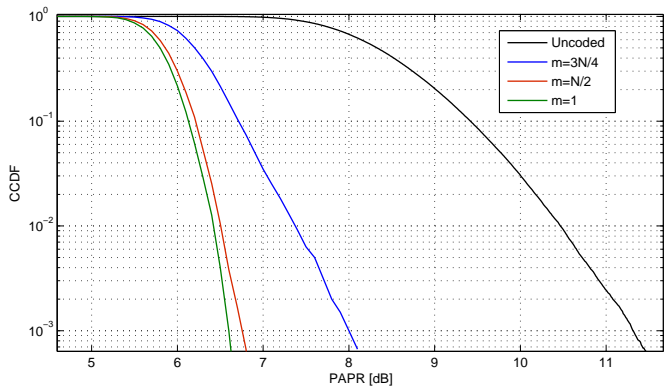}
	\vspace{-0.7cm}
	\caption{PAPR reduction of the refined algorithm for $n=256$, 16-QAM and $q=100$.}
	\label{fig:n256-pruned}
	\vspace{-0.3cm}
\end{figure}

Performance of the refined version of the algorithm for $m=n/2$ and~$3n/4$ is shown in \fig{fig:n64-pruned-bounds} and \fig{fig:n256-pruned} for $n=64$ and~$256$, respectively. The modulation used here is 16-QAM. A remarkable observation is that by using only the second half of the sign changes, i.e., 0.5~b/sym of rate loss, almost the same reduction is achieved. Moreover, by using only the last one fourth of the sign changes, i.e., 0.25~b/sym of rate loss, a reduction of about 3~dB is achievable. This behavior is particularly interesting as it allows for hybrid ideas to exploit this abundance of sign variables. As mentioned before, the rate loss decreases for a higher constellation size.

\section{Conclusion}
\label{sec:conclusion}
A novel sign selection algorithm is proposed for PAPR reduction of OFDM waveform. It belongs to the class of algorithms with high PAPR reduction capability, which comes at the cost of high computational complexity. The proposed algorithm, in its basic form, offers a PAPR reduction of about 4.6 dB for 64 subcarriers and 16-QAM. A slightly higher reduction is achieved for higher number of subcarriers. A refined version of the algorithm uses a subset of the sign variables and offers a rate loss of 0.5 bits per complex data symbol, instead of 1 b/sym of the basic algorithm. 

Mathematical analysis of the algorithm shows that the reduced PAPR can be upper-bounded in a probabilistic way, which offers much tighter bounds in comparison with the worst-case upper bounds which are often not insightful. This is done analytically by using concentration inequalities, and numerically by estimation of initial partial expectations.

Performance of the algorithm for advanced multicarrier waveforms clearly needs to be examined. Although a naive extension of the algorithm to these waveforms is straightforward, a proper and general investigation is beyond the scope of this paper and is left as a topic for further research. 

\section*{Acknowledgements}
This work was supported by German Research Foundation (DFG) under grant WU 598/3-1.

\bibliographystyle{IEEEtran}
\bibliography{biblio}

\end{document}